\documentclass[manuscript]{emulateapj}
\def \kmsec {\rm km~s$^{-1}$\,}

\def \al {\rm et al.\,}
\def \13CO {$^{13}$CO\,}
\def \C18O {C$^{18}$O\,}

\def \Tmb {$T_{\rm mb}$\,}

\def \Msun {$M_{\odot}$\,}

\def \arcmin {$^\prime$\ }

\def \j#1#2{$J$$=#1-#2$\,}

\def \3P1 {$^3$P$_1$ -- $^3$P$_0$\,}

\def \H2 {H$_{2}$\,\,}
\def \nh2 {{\it n}(H$_2$)}

\newcommand{\asec}{$^{\prime \prime}$}

\newcommand{\halpha}{H$\alpha$}                   %H I recombination lines 

\shorttitle{Methanol in the L1551 Circumbinary Torus}
\shortauthors{White et al.}

\begin{document}


\title{Methanol in the L1551 Circumbinary Torus}


\author{Glenn J. White\altaffilmark{1}$^,$\altaffilmark{2}}
\author{C.W.M.Fridlund\altaffilmark{3}}
\author{P.Bergman\altaffilmark{4}}
\author{A.Beardsmore\altaffilmark{5}}
\author{Rene Liseau\altaffilmark{6}}
\author{M.Price\altaffilmark{5}}
\author{R.R.Phillips\altaffilmark{7}}


\altaffiltext{1}{Department of Physics $\&$ Astronomy, The Open University, Walton Hall, Milton Keynes MK6 7AA, England. g.j.white@open.ac.uk}
\altaffiltext{2}{Space Physics Division, The Rutherford Appleton Laboratory, Chilton, Didcot OX11 0QX, England}
\altaffiltext{3}{ESTEC, P.O. Box 299, NL-2200AG Noordwijk, Netherlands. Malcolm.Fridlund@esa.int}
\altaffiltext{4}{Onsala Space Observatory, SE-439 92 Onsala, Sweden. bergman@oso.chalmers.se}
\altaffiltext{5}{University of Kent, Canterbury CT2 7NR, England. mcp2@star.kent.ac.uk, ab75@star.kent.ac.uk}
\altaffiltext{6}{Stockholm Observatory, 106 91 Stockholm, Sweden. rene@astro.su.se}
\altaffiltext{7}{University of Lethbridge, 4401 University Drive, Lethbridge, Alberta T1K 3M4, Canada. robin.phillips@uleth.ca}

\begin{abstract}
We report observations of gaseous methanol in an edge-on torus surrounding the young stellar object L1551 IRS5. The peaks in the torus are separated by $\sim$ 10,000 AU from L1551 IRS5, and contain $\sim$ 0.03 $M_{\rm \earth}$ of cold CH$_3$OH. We infer that the CH$_3$OH abundance increases in the outer part of the torus, probably as a result of methanol evaporation from dust grain surfaces heated by the shock luminosity associated with the shocks associated with the jets of an externally located x-ray source. Any methanol released in such a cold environment will rapidly freeze again, spreading CH$_3$OH throughout the circumbinary torus to nascent dust grains, planitesimals, and primitive bodies. These observations probe the initial chemical conditions of matter infalling onto the disk.
\end{abstract}




%\vspace*{-1mm}
\section{Introduction}
In the earliest stages of their formation, low-mass young stellar objects are embedded in  flattened gaseous envelopes. Surrounding these nascent protostars, planetary systems, primitive bodies and their attendant dust grains will condense from the infalling material. In this paper, we study the distribution of the organic molecule methanol around a protostellar core, as it is an important constituent of the young material in the disk. The L1551 molecular core (distance 140 pc) contains a young binary system (seen as an infrared and radio source L1551 IRS-5) hidden inside a dense envelope providing $\sim$ 150 magnitudes of visual extinction (Stocke \al 1988, White \al 2000, Fridlund \al 2005). A molecular outflow emanates from the center of the disk (Snell \al 1980, Kaifu \al 1984, Fridlund \& White 1989, Rainey \al 1987, Parker \al 1991), with atomic jet(s) associated with an x-ray emitting region being observed to the SW (Fridlund \& Liseau 1998, Favata \al 2002, 2003, Bally \al 2003, Fridlund \al 2005). The protostellar disk is surrounded by a massive (radius $\sim$ 20,000 AU) cool envelope that exhibits both rotational and infall motions (Takakuwa \al 2004, Moriarty-Schieven \al 2006). Fridlund \al (2002) used HCO$^+$, H$^{13}$CO$^+$ and $^{13}$CO \j10 observations to estimate that the mass of the disk is $\sim$ 2.5$\pm$1.5 \Msun.
%\vspace*{-1mm}
\section{Observations}
%% In a manner similar to \objectname authors can provide links to dataset
%% hosted at participating data centers via the \dataset{} command.  The
%% second curly bracket argument is printed in the text while the first
%% parentheses argument serves as the valid data set identifier.  Large
%% lists of data set are best provided in a table (see Table 3 for an example).
%% Valid data set identifiers should be obtained from the data center that
%% is currently hosting the data.
%%
%% Note that AASTeX interprets everything between the curly braces in the 
%% macro as regular text, so any special characters, e.g. "#" or "_," must be 
%% preceded by a backslash. Otherwise, you will get a LaTeX error when you 
%% compile your manuscript.  Special characters do not 
%% need to be escaped in the optional, square-bracket argument.

Observations of CH$_3$OH (96 GHz 2$_k$--1$_k$ and 242 GHz 5$_k$--4$_k$), HCN (\j10), and CO (\j10) were obtained using the Onsala 20m and JCMT 15m telescopes in Feb 2004 and Jan/Apr 2005. The Onsala observations used an single sideband SIS receiver, with a 1600 channel $\times$ 25 kHz autocorrelator backend. The JCMT observations of a higher CH$_3$OH transition (241 GHz 5$_k$--4$_k$) were made using the standard facility RxA2 receiver. All lines were corrected to a common main beam brightness, \Tmb scale. Additional data for CS and C$^{34}$S (\j21) are taken from Fridlund \al (2002).

%\vspace*{-1mm}
\section{Methanol}
%\vspace*{-1mm}
%label{methanol}
Methanol plays a key part in the chemistry leading to the production of biogenic molecules, and is an important constituent of icy grain mantles and comets. Thermally excited gas phase CH$_3$OH has previously been observed from warm inner envelopes surrounding just a few Class 0 protostars, where it is evaporated from warm dust grains heated by the central protostar, or by an accretion shock at the edge of an infalling envelope (Goldsmith \al 1999, Velusamy \al 2002, Maret \al 2005). 

Maps showing the distribution of CH$_3$OH, HCN and CO toward L1551 observed from Onsala are shown in Figure 1, and velocity channel maps for the 96 GHz CH$_3$OH and CS \j21 lines in Figure 2.

\begin{figure}
\epsscale{1.2}
\plotone{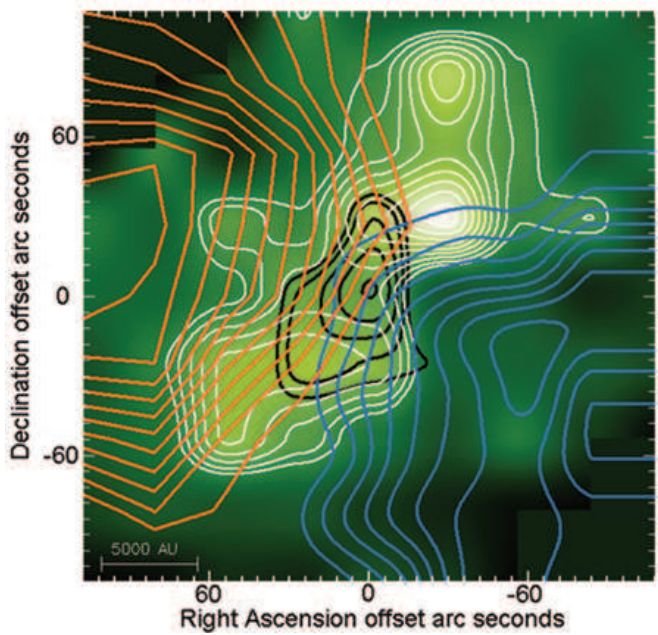}
\caption{CH$_3$OH 2$_{0}$--1$_{0}$($A$) line map (Onsala data at 96.7414 GHz, green image with white contours) superimposed with contours of the HCN J= 1--0 (Onsala data shown as black contours), and (red) and (blue) shifted CO J= 1--0 outflows (shown as red and blue contour lines respectively) from the data shown in Fridlund \al (2002). West is on the left, north at the top. The bottom left bar shows the size scale, for a distance of 140 pc.}
\label{compositemap}
\end{figure}

\begin{figure}
\includegraphics[scale=.47,angle=270]{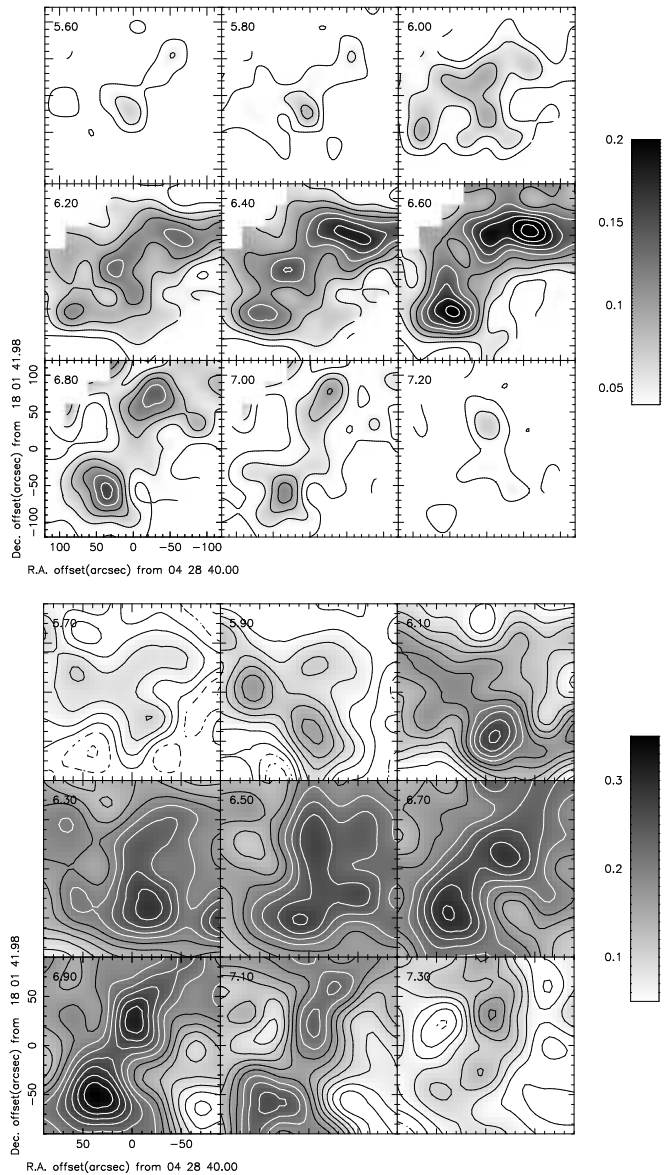}
\break
\includegraphics[scale=.47,angle=270]{f2b.eps}
\caption{[upper] Channel maps in the 2$_0$--1$_0$($A$) CH$_3$OH line.  The lowest contour level is 0.15 K km s$^{-1}$ and contour intervals are at 0.1 K km s$^{-1}$, [lower] Channel maps in the CS \j21 line with the lowest contour at 0.375 K \kmsec channel, and subsequent contour intervals increasing by 0.125 K \kmsec. The central (0,0) position of the maps and spectra is that of L1551 IRS5 at RA(2000) = 04$^h$ 31$^m$ 34$^s$.13, Dec(2000) = 18$^{\circ}$ 08{\arcmin} 04$^{\prime\prime}$.95.
\label{methanolchannelmaps}}
\end{figure}

The methanol has three peaks located $\sim$ 6500--12000 AU from IRS5 along a SE-NW line. The high density tracer, HCN is centrally peaked and marginally resolved, located between the 2 inner methanol peaks. CO emission, which traces low density material, streams orthogonally away to the NE and SW. The methanol appears as an edge-on torus surrounding the HCN core, and is oriented orthogonally to the CO outflow. The velocity shift between the two CH$_3$OH peaks is 0.07 km s$^{-1}$ -- which ${\it if}$ indicative of rotation, would imply a rotational period $\sim$ 6.5 $\times$ 10$^5$ yr. Spectra toward L1551 IRS-5 and at the two main methanol peaks are shown in Figure 3, along with the model fits described next.

\begin{figure}
\epsscale{1.0}
\includegraphics[scale=.47]{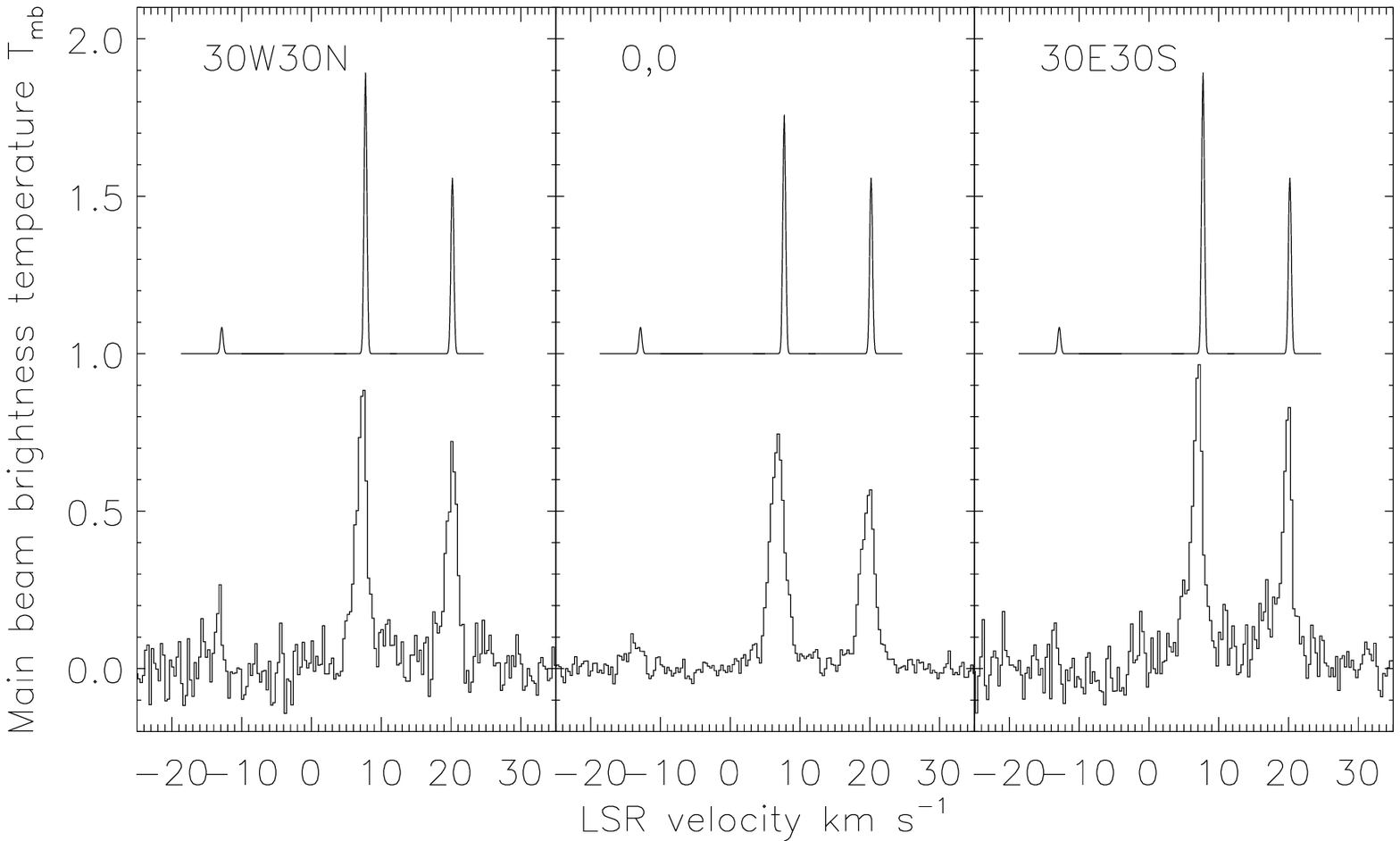}
\break
\includegraphics[scale=.47]{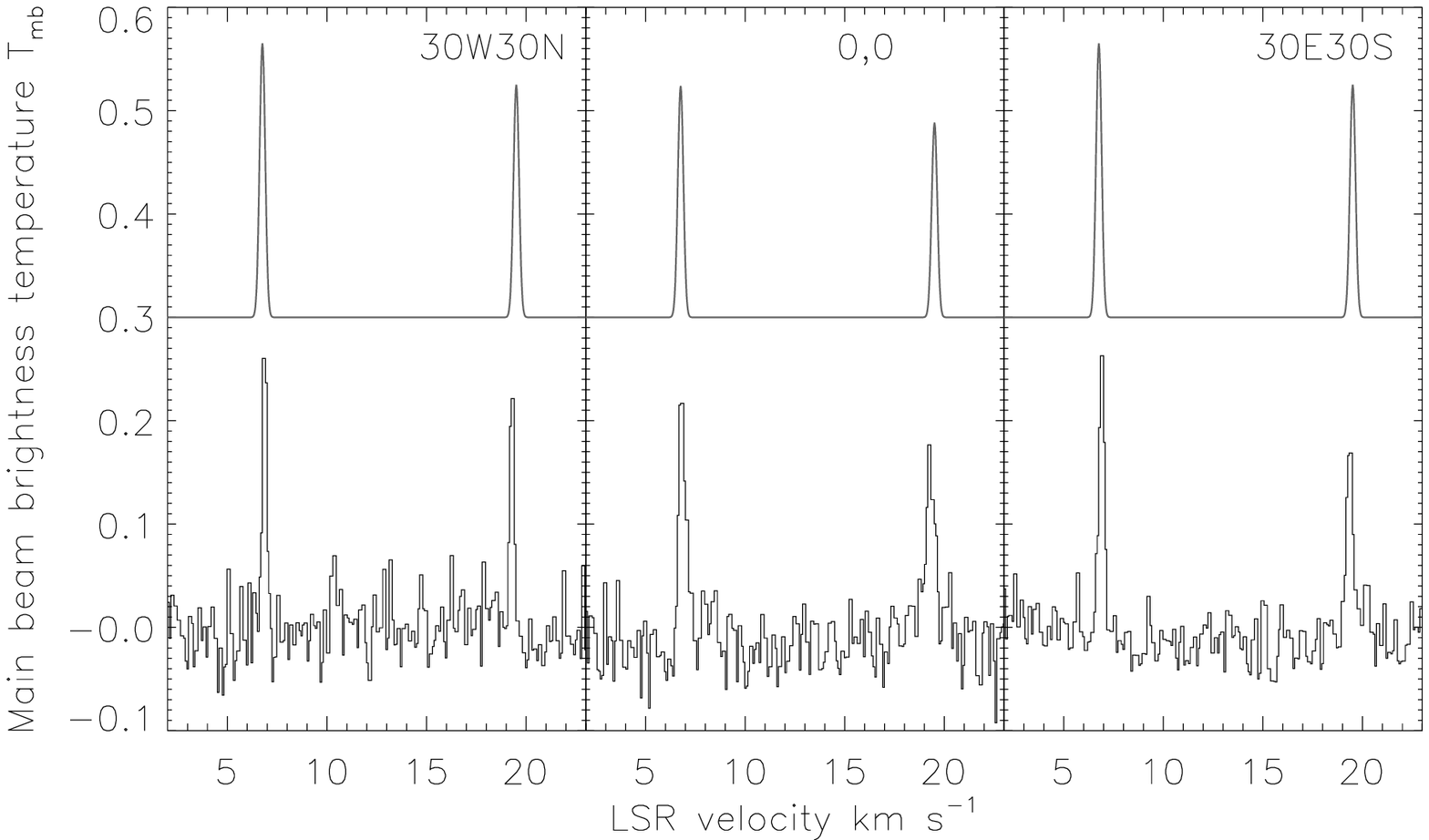}
\caption{CH$_3$OH data and model fits for the (left to right) 2$_{-1}$--1$_{-1}$($E$), 2$_{0}$--1$_{0}$($A$) and 2$_{0}$--1$_{0}$($E$) lines at 96.7394, 96.7414 and 96.7446 GHz, and the 5$_{-1}$--4$_{-1}$($E$), and 5$_{0}$--4$_{0}$($A$), lines at 241.7672 and 241.7914 GHz respectively toward the central IRS5 position, and at the two peaks in the 96 GHz CH$_3$OH line. The frequency scales assume v$_{lsr}$ = 6.5 \kmsec (Fridlund \al 2002)}
\label{spec96}
\end{figure}

We have modeled the CH$_3$OH line intensities using an accelerated lambda iteration radiative transfer technique for a spherical-core envelope (Phillips $\&$ Little 2000), with an assumed core radius of 5000 AU; envelope radius of 12,500 AU; and turbulent velocity of 0.45 km s$^{-1}$ (from the H$^{13}$CO$^+$ linewidth - Fridlund \al 2002). The $A$ and $E$ symmetry states of CH$_3$OH were calculated separately using collisional rates, energy levels and Einstein rates from M${\rm \ddot{u}}$ller \al (2001) and Pottage \al (2004), and convolved to the Onsala and JCMT beamwidths. A number of models were run spanning a range of temperature, density and abundance gradients to simultaneously match both the relative and absolute spectral line intensities, and the observed morphology of the torus. The observations unfortunately do not provide strong constraints on the CH$_3$OH kinetic temperature, and so the temperature and density gradients used in the model were chosen to match those of White \al 2000 derived from dust continuum measurements.

Using this model the best fit suggests a central hydrogen gas density (at the position of L1551 IRS-5) ${\it n}$(H$_2$) = 7.5 x 10$^5$ cm$^{-3}$ and a methanol abundance $X$ = 8$\times$10$^{-10}$ relative to H$_2$. Since the temperature and density are known to decline away from L1551 IRS5 (see Figure 4a, which is derived from White \al 2000), the observed torus structure then requires that the methanol abundance increases to $\sim$ 4 x 10$^{-9}$ at the inner edge of the methanol ring, and to 10$^{-8}$ in the outer envelope, as shown in Figure 4b.

\begin{figure}
%\epsscale{0.4}
%\plottwo{f4a.eps}{f4b.eps}
%\plotone{f4.eps}
%\includegraphics[angle=-90,width=\textwidth]{f4.eps}
\epsscale{1.0}
%\plottwo{f4a.eps}{f4b.eps}
\includegraphics[scale=.47,angle=0]{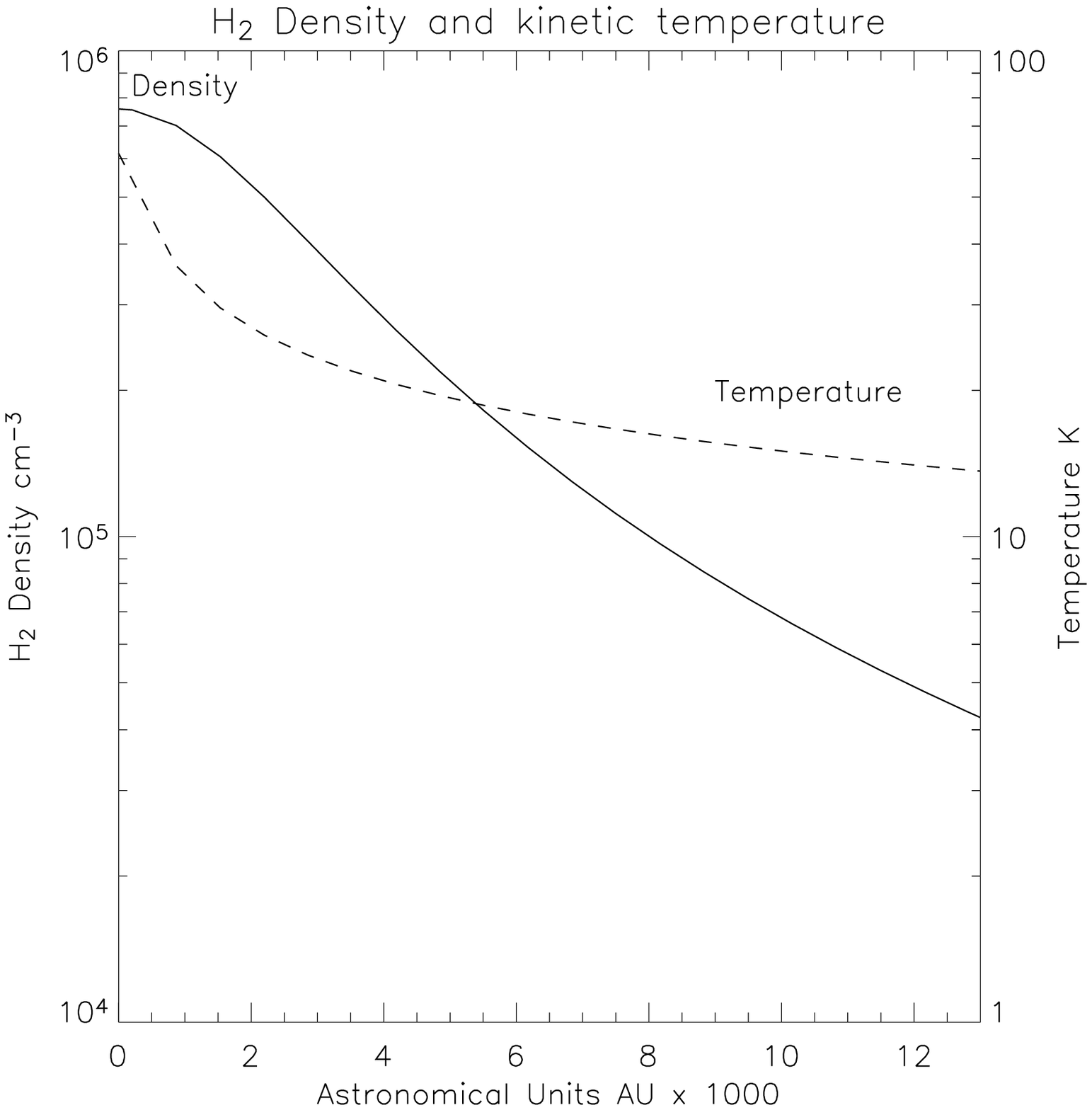}
\includegraphics[scale=.47,angle=0]{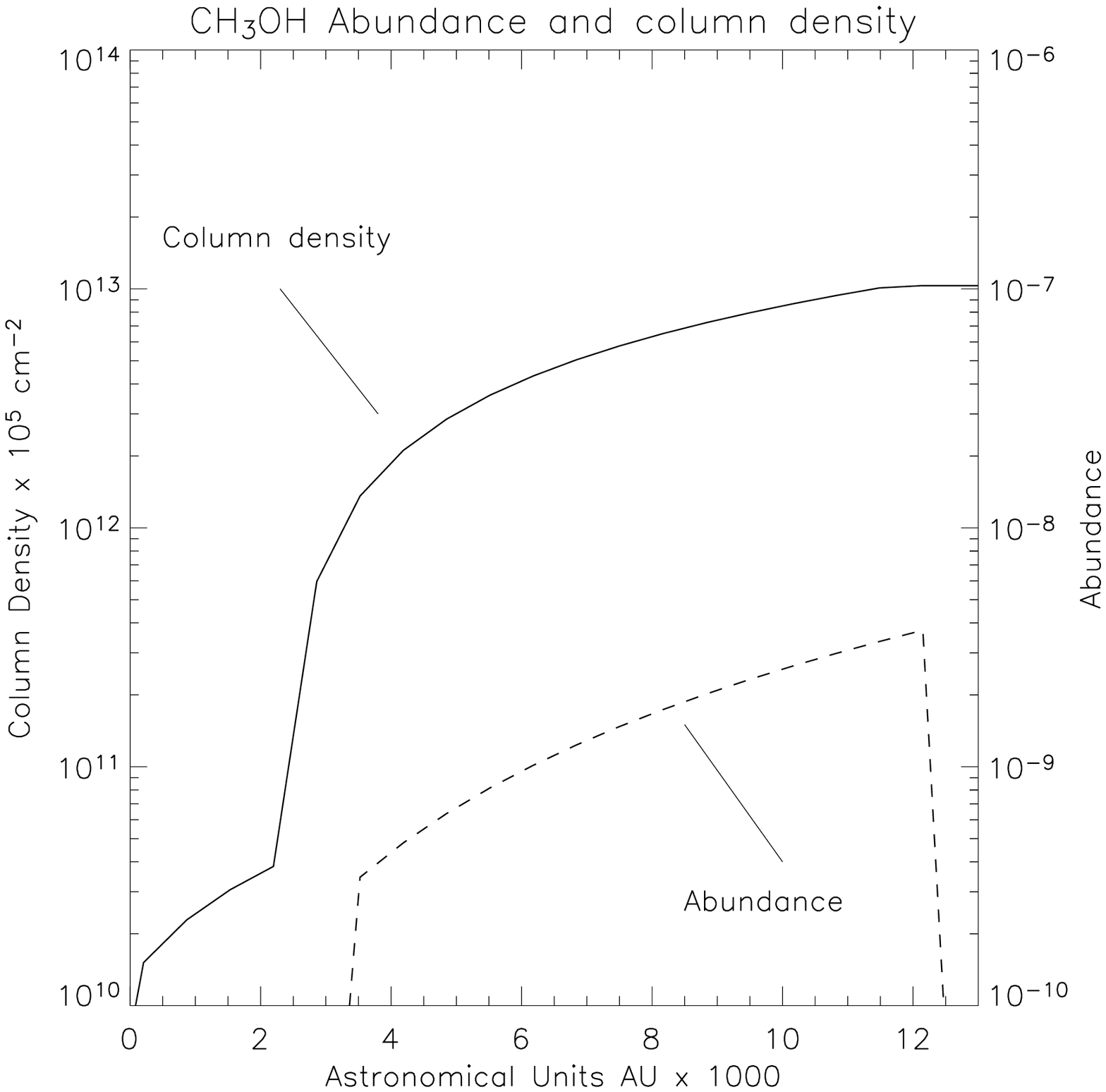}
\caption{(a) Temperature and density from White \al (2000) used in the CH$_3$OH model, (b) abundance and column density estimated from the best fit CH$_3$OH model
\label{tandrho}}
\end{figure}

The abundance increase is needed since H$_2$ density is $\sim$ 20 times lower at the edge of the CH$_3$OH peaks than towards the source centre at L1551 IRS-5. In this best fit model, the torus contains 0.03 $M_{\rm \earth}$ of CH$_3$OH. Models run with CH$_3$OH kinetic temperatures in the torus $\geq$20K failed to reproduce the observed disk structure whilst simultaneously matching the absolute and relative line temperatures. Since the gas and dust should be well thermalised at the estimated densities, it is reasonable to assume that the kinetic temperature of the CH$_3$OH will be close to the dust temperature (13K from White \al 2000). The models underestimate the observed linewidth toward IRS-5, which we believe is affected by additional dynamics and/or turbulence/outflow activity near L1551 IRS5 which is not modeled here.

%\vspace*{-1mm}
\section{Other species in the circumbinary torus}

The integrated HCN, CS, C$^{34}$S, HCO$^+$ and H$^{13}$CO$^+$ distributions peak close to L1551 IRS-5 (Fridlund \al 2002), whereas the CH$_3$OH map appears as an edge-on torus surrounding the HCN core. The CS and HCO$^+$ channel maps show a faint signature of this torus structure, but at low levels not contributing significantly to their integrated emission maps (Fridlund et al ${\it in ~preparation}$).

Van der Tak \al (2000) suggest that gas phase production of CH$_3$OH at temperatures $\leq$100 K follows the radiative association of H$_2$O and CH$^{+}_{3}$. However, the relative inefficiency of the process would result in low abundances $\sim$ 10$^{-11}$ relative to H$_2$. Recent storage ring measurements (Geppert \al 2006) of the dissociative recombination of CH$_3$OH$_2$$^+$ with electrons indicate extensive C--O bond fragmentation, emphasising the difficulty for ion-molecule reactions to form CH$_3$OH, which will preferentially be destroyed via an H$^{+}_{3}$ + CH$_3$OH $\longrightarrow$ CH$_3$OH$^{+}_{2}$ + e$^+$ reaction. It is increasingly difficult to reconcile observed CH$_3$OH abundances with ion-molecule chemical production. Solid phase reactions, including cosmic ray irradiation processes therefore likely are important for CH$_3$OH formation on grain surfaces (Wada \al 2006). Observations in dark clouds, where grain chemistry operates, indicate higher abundances of $\sim$ 10$^{-9}$ (Takakuwa \al 1998, Turner 1998), and toward hot cores, where CH$_3$OH is evaporated off grain surfaces, abundances $\ga$ 10$^{-7}$ (Blake \al 1987) have been reported. Grain temperatures would need to be $\ga$ 120 K for methanol to be efficiently released from icy or clathrate mixtures on dust grain surfaces (Blake \al 1987, Sandford $\&$ Allamandola 1993).

%\vspace*{-1mm}
\subsection{Heating of grains}
Our model suggests that gas phase methanol in the envelope is sufficiently cold to rapidly freeze back onto dust grains and solid surfaces, which will occur on a timescale (${\it t} \sim$ 3 $\times$ 10$^9$/$n(H)$ $\sim $ 4000 years, where ${\it n(H)}$ is the hydrogen density cm$^{-3}$). Shocks and direct heating, could re-evaporate the icy mantles, although the liberated CH$_3$OH will rapidly cool again. The mid-plane dust temperature, T$_{dust}$ due to L1551 IRS5 will be (Fridlund \al 2002):
%\vspace*{-1mm}
\begin{equation}
{\rm T}_{dust}  \approx 38\;\left( {\frac{r}{{100\;{\rm AU}}}} \right)^{ - 0.4} \left( {\frac{L}{{L_{\odot} }}} \right)^{0.2} \quad {\rm K}\label{colden}
\end{equation}

At the CH$_3$OH peaks, T$_{dust}$ $\sim$ 13 K (White \al 2000), which is inadequate for CH$_3$OH to be evaporated. Methanol has been observed toward the L1157 molecular disk (Velusamy \al 2002), and higher mass protostellar cores (van der Tak \al 2000), where direct heating could directly evaporate the CH$_3$OH - but it does not appear that this mechanism could drive off significant amounts of CH$_3$OH at distance of the L1551 CH$_3$OH peaks. 

An alternative source of heating is suggested by the presence of an Herbig-Haro jet (HH154) emanating from L1551 IRS5, and terminating in a working surface -- presently located $\sim$ 15\asec~away from the IRS5 binary (Fridlund \al 2005). This jet is one of the few X-ray sources associated with low mass stellar jets, and has 4 separate shock interactions -- two detected in X-rays (Favata \al 2002, 2006, Bally \al 2003), and two seen in [OIII] 5007\AA ~and \halpha ~(Fridlund \al 2005). Assuming a clear line of sight to the surface of the flared disk/envelope/torus surface, the temperatures of icy grains heated by the jet luminosity would be (Hollenbach \& McKee, 1979):
 %\vspace*{-1mm}
 \begin{equation}
 {\rm T}_{grain} = 47~\left[\frac{J_{UV}}{a_\mu \times C_1}\right] ~~{\rm K}
 \end{equation}
for ice, where C$_1$~is 2, the grain radius, a$_{\mu}$ is assumed to be 0.15$\mu$m (Brown, 1990) and the mean intensity is $J_{UV}$.  The shock luminosity is then: 
%\vspace*{-1mm}
\begin{equation}
F_S = 5.8 \times 10^{-4} n_0 v_{s7}^3 ~~{\rm ~erg ~cm^{-2} ~s^{-1}} 
\end{equation}
% \vspace*{-2mm}
The projected distance of the methanol peaks from the shocks is $\approx$~6000 AU. Assuming this flux is dominated by X-rays, EUV and FUV emission (Hollenbach \& McKee 1979; Hartigan \al 1987) ${\rm T}_{grain} $ will reach 100--150 K - which is adequate to liberate CH$_3$OH. The grains would however rapidly cool again, with the evaporated gas becoming thermally well coupled with the grains. The infall of material would naturally move grains initially at the edge of the torus from the zone exposed to x-rays and EUV/VUV photons to cooler shielded regions of the torus.

The gas phase CH$_3$OH abundance in L1551's outer molecular envelope will depend on the balance between several processes including: ${\it a}$) freeze-out onto dust grains; ${\it b}$) destruction by ion-molecule reactions, since enhanced ion densities will be induced by the x-ray irradiation forming CH$_3$OH on a similar timescale to grain surface reactions (Wada \al 2006); and ${\it c}$) evaporation or release from dust grains following grain heating, or vapourisation by an accretion shock (Velusamy, Langer $\&$ Goldsmith 2002).

To test the importance of ${\it b}$) above, a chemical model was kindly run for us by Dr S. Viti  (Lintott $\&$ Viti 2006 $in ~preparation$) for a molecular cloud having solar metallicity and a high (100 x standard) ionization rate, with the same temperature and density to those we infer. This suggests that a high methanol abundance could survive against ion-molecule reactions for at least 10$^4$ yrs ${\it i.e.}$ similar to or greater than the freeze-out timescale. For ${\it c}$), we favor release from heated grains over the accretion shock scenario on the basis of the narrow CH$_3$OH linewidths that are observed in the outer L1551 molecular envelope ($\sim$ 0.6 \kmsec compared to several \kmsec in the L1157 inner core), and the distance of the CH$_3$OH peaks from L1551 IRS-5. There is no compelling evidence to suggest the CH$_3$OH abundance in the outer L1551 torus is significantly influenced by an accretion shock: the narrow CH$_3$OH line center lies within 0.1 \kmsec of the systemic velocity inferred from many other lines ($\sim$ 6.5 \kmsec -- Fridlund \al 2002).

%\vspace*{-2mm}
\subsection{Discussion and Conclusions}
Star formation occurs in the dense cores of molecular clouds. Assuming that comet formation occurs as an adjunct to this process, information about the primordial chemical abundances should be locked into the the dust grains and comets, which are the main repositories of primitive material left over from the solar protostellar disc. The molecular inventory of protostellar envelope material has been discussed by White \al (2003), with CH$_3$OH commonly being found in comets (Mumma \al 2003), along with other organic molecules. Our detection of methanol in a region at a similar distance from L1551 IRS5 as that of the Oort Cloud from our own Sun, suggests that the CH$_3$OH may be able to accrete to and spread throughout the outer L1551 protostellar envelope, seeding dust grains and primitive bodies with chemical precursor material that could contribute to the synthesis of more complex molecules. This detection of methanol in the L1551 circumbinary torus provides a first glimpse into the initial chemical conditions of matter infalling onto protostellar disks, and future determinations of the abundances of other molecules in the X-ray/EUV illuminated zone could therefore provide a interesting test of the "interstellar" versus "nebular" aspects of cometary and planetesimal chemistry.\\
\\
In summary: ${\it a}$) we have detected a massive (0.03 $M_{\rm \earth}$), and most likely cold ($\leq$ 20K -- based on dust continuum measurements), toroidal shaped ring of methanol surrounding the L1551 IRS5 protobinary system, ${\it b}$) we propose a new mechanism where shock luminosity heating by the jets associated with an externally located x-ray source -- these have a direct line of sight to the torus surface could raise the dust temperature sufficiently to liberate CH$_3$OH, ${\it c}$) we suggest that the luminosity from the jet/x-ray source located just above the L1551 torus is able to heat the dust grains sufficiently to liberate a substantial amount of gaseous CH$_3$OH in the outer disk of L1551, and ${\it d}$) we speculate that the methanol released will rapidly freeze back onto solid material in the outer envelope, spreading CH$_3$OH to the surfaces of dust grains and primitive bodies - and potentially modifying the primitive surfaces of solid material in L1551's outer molecular envelope. The observations give indications of the initial chemical conditions of matter infalling onto the disks.\\


%% Included in this acknowledgments section are examples of the
%% AASTeX hypertext markup commands. Use \url without the optional [HREF]
%% argument when you want to print the url directly in the text. Otherwise,
%% use either \url or \anchor, with the HREF as the first argument and the
%% text to be printed in the second.
%\vspace*{-8.5mm}
\acknowledgments
We acknowledge discussions with B. Davidson, G. Moriarty-Schieven, V. Ossenkopf, R. Nelson, H. Rickman, F. van der Tak, M. Walmsley, D. Williams and S. Viti, and thank D. Clements and H. Butner for help in obtaining the JCMT service observations, and an anonymous referee for helpful comments.

\clearpage

%% the file, uncomment the \plotone, \plottwo, and
%% If you need a layout that cannot be achieved with \plotone or
%% \plottwo, you can invoke the graphicx package directly with the
%% \includegraphics command or use \plotfiddle. For more information,
%% please see the tutorial on "Using Electronic Art with AASTeX" in the
%% documentation section at the AASTeX Web site,
%% http://www.journals.uchicago.edu/AAS/AASTeX.
%%
%% The examples below also include sample markup for submission of
%% supplemental electronic materials. As always, be sure to check
%% the instructions to authors for the journal you are submitting to
%% for specific submissions guidelines as they vary from

%% This example uses \plotone to include an EPS file scaled to
%% 80% of its natural size with \epsscale. Its caption
%% has been written to indicate that additional figure parts will be
%% available in the electronic journal.

%% Here we use \plottwo to present two versions of the same figure,
%% one in black and white for print the other in RGB color
%% for online presentation. Note that the caption indicates
%% that a color version of the figure will be available online.
%%

%%
%%
%%
%%
%%

\end{document}